\documentclass[german,english]{bvm2019}

\title{Navigierte Interventionen im Kopf- und Halsbereich: Standardisiertes Assessment eines neuen, handlichen Feldgenerators}
\titlerunning{Standardisiertes Assessment eines neuen, handlichen Feldgenerators}

\author{Benjamin~J.~Mittmann$^{1,3}$, Alexander~Seitel$^3$, Lena~Maier-Hein$^3$,  Alfred~M.~Franz$^{2,3}$}

\authorrunning{Mittmann et al.}

\institute{
$^1$ Institut für Medizintechnik und Mechatronik, Hochschule Ulm\\
$^2$ Institut für Informatik, Hochschule Ulm\\
$^3$Abteilung Computer-assistierte Medizinische Interventionen, DKFZ Heidelberg
}

\email{mittmann@mail.hs-ulm.de}

\begin{document}

%
\selectlanguage{german}

\maketitle

\begin{abstract}
Elektromagnetische (EM) Trackingsysteme verwenden zur Lokalisation von OP-Instrumenten am Eingriffsort ein EM Feld, das von einem Feldgenerator (FG) erzeugt wird. Üblicherweise sind die FG umso größer, je höher die Reichweite ihres Trackingvolumens ist. Der kürzlich von der Firma NDI (Northern Digital Inc., Waterloo, ON, Canada) vorgestellte \textit{Planar 10-11 FG} vereint erstmals eine kompakte Bauweise und ein dazu verhältnismäßig großes Trackingvolumen. Mit einem standardisierten Messprotokoll wurde der FG auf seine Robustheit gegenüber externen Störquellen und seine Genauigkeit geprüft. Die mittlere Positionsgenauigkeit beträgt 0,59~mm (\textit{Standard-Setup}) bei einem mittleren Jitter von 0,26~mm. Der mittlere Orientierungsfehler fällt mit 0,10° sehr gering aus. Der höchste durch ein Metall verursachte Positionsfehler (4,82~mm) wird von Stahl SST 303 hervorgerufen. Bei Stahl SST 416 ist der Positionsfehler (0,10~mm) am geringsten. Im Vergleich zu zwei anderen FG von NDI erreicht der \textit{Planar 10-11 FG} tendenziell bessere Genauigkeitsergebnisse. Wegen seiner Kompaktheit und der damit verbundenen mobilen Einsatzfähigkeit könnte der FG daher dazu beitragen, den Gebrauch von \mbox{EM Trackingsystemen in der Klinik zu steigern.}
\end{abstract}

\section{Einleitung}
Bei medizinischen Interventionen spielt die Beachtung angrenzender Risikostrukturen allgemein eine bedeutende Rolle. Durch den Einsatz sogenannter intraoperativer Trackingsysteme (IOT) kann die Wahrscheinlichkeit, eine Risikostruktur ungewollt zu verletzen, minimiert werden. Sie ermöglichen eine genaue Lokalisation von OP-Instrumenten am Eingriffsort und bieten dem Operateur visuelle Hilfestellungen durch Augmented-Reality-Darstellungen. In der \mbox{Neurochirurgie,} einem Gebiet, das besondere Präzision vom Operateur verlangt, kommen IOT standardmäßig zum Einsatz~\cite{2639-01}. In vielen anderen Fachbereichen hingegen hat sich der Routineeinsatz von IOT im Klinikalltag bislang nicht durchgesetzt. 

Wie Studien der vergangenen Jahre belegen, könnten insbesondere minimalinvasive Eingriffe im Hals- und Kopfbereich künftig von IOT profitieren. Beispiele hierfür sind perkutane Punktionen wie die Punktion von Knoten innerhalb der Schilddrüse~\cite{2639-02}, endoskopische Interventionen wie die Endoskop-geführte Nasennebenhöhlenchirurgie~\cite{2639-03} oder Kathetereingriffe wie die Thrombektomie zur Behandlung eines akuten Schlaganfalls. In der aktuellen Forschung zur Thrombektomie wird darauf verwiesen, dass ein Tracking des Katheters helfen könnte, die Reperfusion der Arterien früher zu ermöglichen~\cite{2639-04}.

Die zur Navigation von OP-Instrumenten notwendige Positionsbestimmung der Instrumente erfolgt in der Regel durch ein optisches System (Optisches Tracking) oder durch ein elektromagnetisches (EM) System~\cite{2639-05}. Optische Tra-ckingsysteme bieten den Vorteil hoher Positionsgenauigkeit. Allerdings ist zur Positionsbestimmung eine durchgängige Sichtlinie (Line-Of-Sight) des optischen Trackers zum getrackten Objekt notwendig. EM Trackingsysteme benötigen keine Line-Of-Sight und können daher potentiell bei perkutanen oder endoskopischen Eingriffen eingesetzt werden. Sie weisen allerdings eine schlechtere Posi-tionsgenauigkeit auf verglichen mit den optischen Systemen~\cite{2639-06}. Für elektromagnetische Trackingsysteme wird ein Feldgenerator (FG) benötigt, der im Raum ein EM Feld erzeugt, in dem EM Sensoren lokalisiert werden können. Das Raumvolumen, in dem die Lokalisation von EM Sensoren möglich ist, bezeichnet man als Trackingvolumen. Üblicherweise nehmen die Größenabmessungen der FG mit der Größe und Reichweite des Trackingvolumens zu. Je größer der FG ist, umso höher ist der Installationsaufwand in der Klinik und umso höher sind die Installationskosten. Dieser Sachverhalt stellt hinsichtlich der praktischen Anwendbarkeit von EM Trackingsystemen im Klinikalltag bis heute ein Problem dar. Ein anderes bekanntes Problem von EM Trackingsystemen ist die mangelnde Robustheit gegenüber externen Störquellen, die das EM Feld des FG verzerren. Eine nahe Platzierung des FG am Zielbereich kann aber die Robustheit des Systems steigern~\cite{2639-07}. Solch eine nahe Platzierung ist häufig jedoch nur mit FG kompakter Bauweise realisierbar.

\begin{figure}[b]
\centering
\includegraphics[width=340pt]{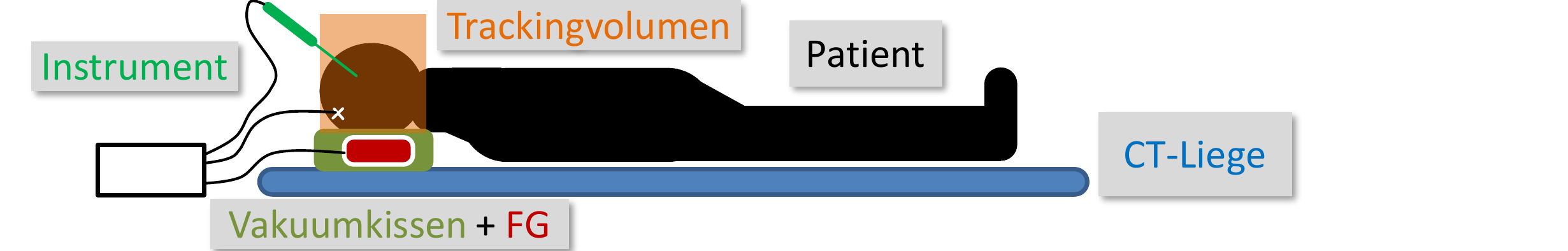}
\caption{\footnotesize Einfaches klinisches Setup für Interventionen im Kopf- und Halsbereich: Der FG ist in ein Vakuumkissen integriert und unterhalb des Patientenkopfes positioniert.}
\label{2639-fig:Setup}
\end{figure}

Kürzlich stellte die Firma NDI den neuen \textit{Planar 10-11 FG} vor. Er vereint erstmals eine kompakte Bauweise (112~mm x 97~mm x 31~mm) und ein dazu verhältnismäßig großes, zylinderförmiges Trackingvolumen (Durchmesser: 340~mm, Höhe: 340~mm). Durch die damit verbundene mobile Einsatzfähigkeit könnte der FG neue Bereiche im klinischen Einsatzgebiet der EM Trackingtechnologie eröffnen. Wie in Abb.~\ref{2639-fig:Setup} skizziert ist, ließe sich der FG z.B. in ein Vakuumkissen integrieren, auf das der Patient seinen Kopf oder seinen Nacken legt. Durch den geringen Abstand des FG zum Kopf befindet sich der gesamte Kopf des Patienten im Trackingvolumen des FG.

Es soll nun untersucht werden, ob der \textit{Planar 10-11 FG} trotz seines relativ großen Trackingvolumens eine vergleichbar hohe Trackinggenauigkeit aufweist, wie andere FG mit ähnlich großen Trackingvolumina. Von Interesse ist dabei auch die Robustheit des FG in Bezug auf Metalle medizinischer Instrumente, da sie die Messgenauigkeit negativ beeinflussen können.

\section{Methoden}
\begin{figure}[b]
\centering
\includegraphics[width=340pt]{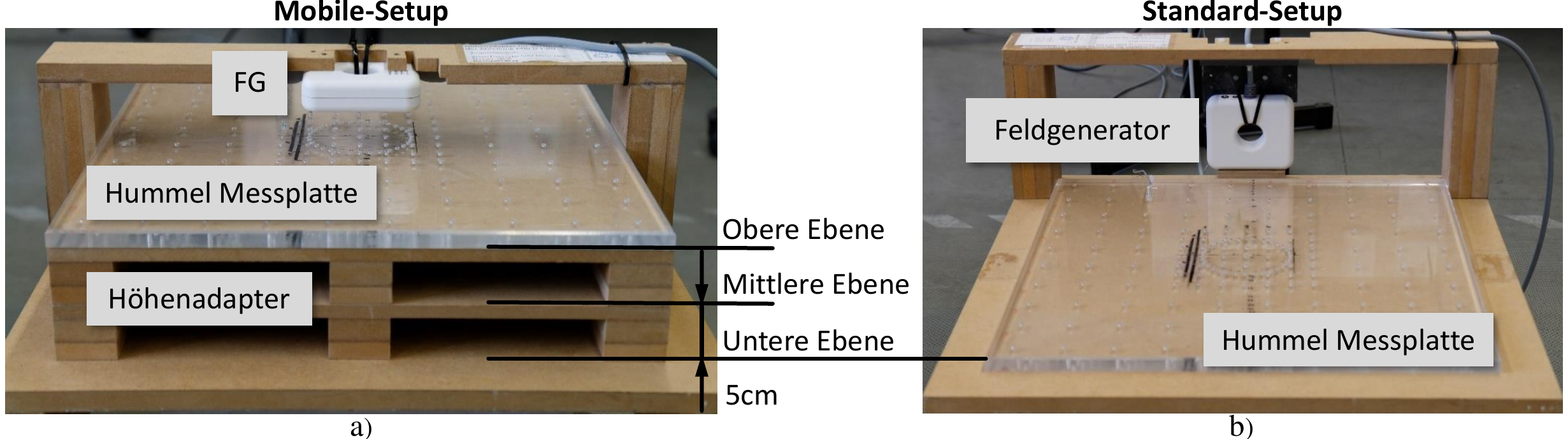}
\caption{\label{2639-fig:Versuchsaufbau}\footnotesize a) Versuchsaufbau beim \textit{Mobile-Setup} für Messungen der oberen Ebene. Der FG ist mittig über der Messplatte montiert. b) Versuchsaufbau beim \textit{Standard-Setup} für Messungen der unteren Ebene. Der FG ist seitlich von der Messplatte befestigt.}
\end{figure}

\begin{figure}[t]
\centering
\caption{\label{2639-fig:HummelPhantom}\footnotesize a) Messplatte nach Hummel et al.~\cite{2639-08}. Eingezeichnet sind die zwei Bereiche, in denen der EM Sensor mit Hilfe der Sensorhalterung für die Positionsmessungen platziert wurde im Abstand von je 5cm. Für die Rotationsmessungen im \textit{Standard-Setup} wurde der FG  seitlich am oberen Rand des 3x4-Gitters platziert. b) Sensormontage an der Sensorhalterung für die Positions- und Rotationsmessungen. Der abgebildete Kreis umfasst 32 Messpositionen mit einem Winkelabstand von je 11,25°. c) Mit Hilfe einer hölzernen Halterung werden 4 verschiedene Metalle nacheinander in den Höhen \textit{H1 - H5} im EM Feld des FG platziert und die Position des Sensors ermittelt.}
\vspace*{0.3cm}
\includegraphics[width=340pt]{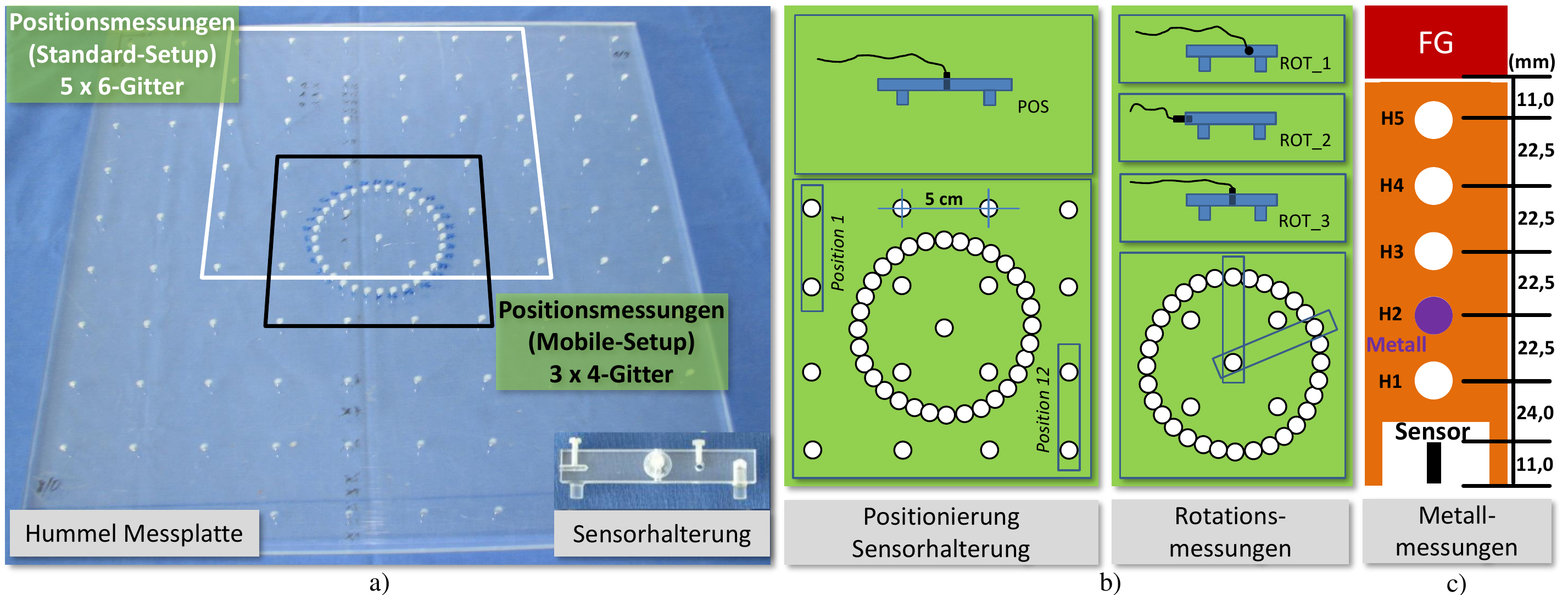}
\vspace*{-0.3cm}
\end{figure}

Zur messtechnischen Bewertung des \textit{Planar 10-11 FG} wurde das \mbox{standardisierte} Messprotokoll von Hummel et al.~\cite{2639-08} verwendet. Um die Vergleichbarkeit der \mbox{Messergebnisse} mit früheren Messungen anderer FG zu gewährleisten, orientierten wir uns an den in~\cite{2639-07} geschilderten Versuchsaufbauten -- dem \textit{Standard-Setup}, bei dem der FG seitlich von der Hummel-Messplatte befestigt ist, und dem \textit{Mobile-Setup}, bei dem der FG mittig über der Messplatte platziert wird (Abb.~\ref{2639-fig:Versuchsaufbau}).
In beiden Setups erfolgten sowohl Positions- als auch Rotationsmessungen. Hierzu wurde der \textit{6DOF Cable Tool; 2,5~mm x 11~mm}-Sensor von NDI an definierten Positionen auf der Messplatte platziert wie in Abb.~\ref{2639-fig:HummelPhantom} veranschaulicht. Für jede Messposition wurden 150 Messwerte innerhalb von 10 Sekunden mit einer Updaterate von 15 Hz aufgezeichnet. Die Wurzel aus den mittleren quadratischen Abweichungen (RMSE) der 150 Messwerte wurde als Jitter definiert. Die Positionsgenauigkeit ermittelten wir mit Hilfe des Mittelwerts der betragsmäßigen Differenzen zwischen den berechneten mittleren euklidischen \mbox{Distanzen} und der real vorliegenden Distanz (5~cm) direkt benachbarter Sensorpositionen innerhalb des 3x4-Gitters beim \textit{Mobile-Setup} bzw. des 5x6-Gitters beim \textit{Standard-Setup}. Die Wahl der Gittergröße erfolgte beim \textit{Standard-Setup} entsprechend der Größe des Trackingvolumens. Beim \textit{Mobile-Setup} orientierten wir uns an der Gitter\-größe vorheriger Studien zum \textit{Compact FG} von NDI~\cite{2639-07}. Die \mbox{Positionsmessungen} erfolgten für beide Setups auf drei verschiedenen Ebenen (unten, mitte, oben; Ebenenabstand: 5~cm). Den Orientierungsfehler berechneten wir anhand der Winkeldifferenz zwischen Paaren von 32 gemessenen Orientierungen und der bekannten relativen Sensorrotation in Höhe von 11,25°. Im \textit{Mobile-Setup} erfolgten in gleicher Weise wie in~\cite{2639-08} Metallmessungen mit vier verschiedenen Metallen: Stahl SST 303, Stahl SST 416, Bronze und Aluminium. Dabei wurden die \mbox{Metalle} nacheinander in fünf definierten Höhen \textit{H1} - \textit{H5} im EM Feld des FG platziert und die Position des fixierten EM Sensors bestimmt (Abb.~\ref{2639-fig:HummelPhantom}c). Anhand der Abweichung zur Referenzposition des Sensors konnte dann der Einfluss des Metalls auf die Messgenauigkeit bestimmt werden.

Zur Untersuchung eines sehr kleinen Mikrosensors, der in Instrumente wie Nadeln integriert werden kann, erfolgten mit dem \textit{Micro 6DOF Sensor} (0,8~mm x 9,0~mm) von NDI zum Vergleich auf der mittleren Ebene im \textit{Standard-Setup} Positions- und Rotationsmessungen entsprechend Abb.~\ref{2639-fig:HummelPhantom}.

\section{Ergebnisse}
\begin{figure}[t]
\centering
\caption{\label{2639-fig:Diagramme} \footnotesize Ergebnisse zum \textit{6DOF Cable Tool}-Sensor: a) Mittlere Jitter im \textit{Mobile-} und \textit{Standard-Setup}. b) Positionsfehler dargestellt als Box-Plots. Die Rauten zeigen den Mittelwert, die Whisker den min. bzw. max. Fehlerwert. Der Positionsfehler ist definiert als Mittelwert der \textit{n} Differenzen zwischen den gemessenen Distanzen und der bekannten 5~cm Referenzdistanz mit n = 17 (\textit{Mobile-Setup}) bzw. n =  49 (\textit{Standard-Setup}). c)~Positionsfehler durch metallische Zylinder zwischen FG und EM Sensor. }
\vspace*{0.3cm}
\includegraphics[width=340pt]{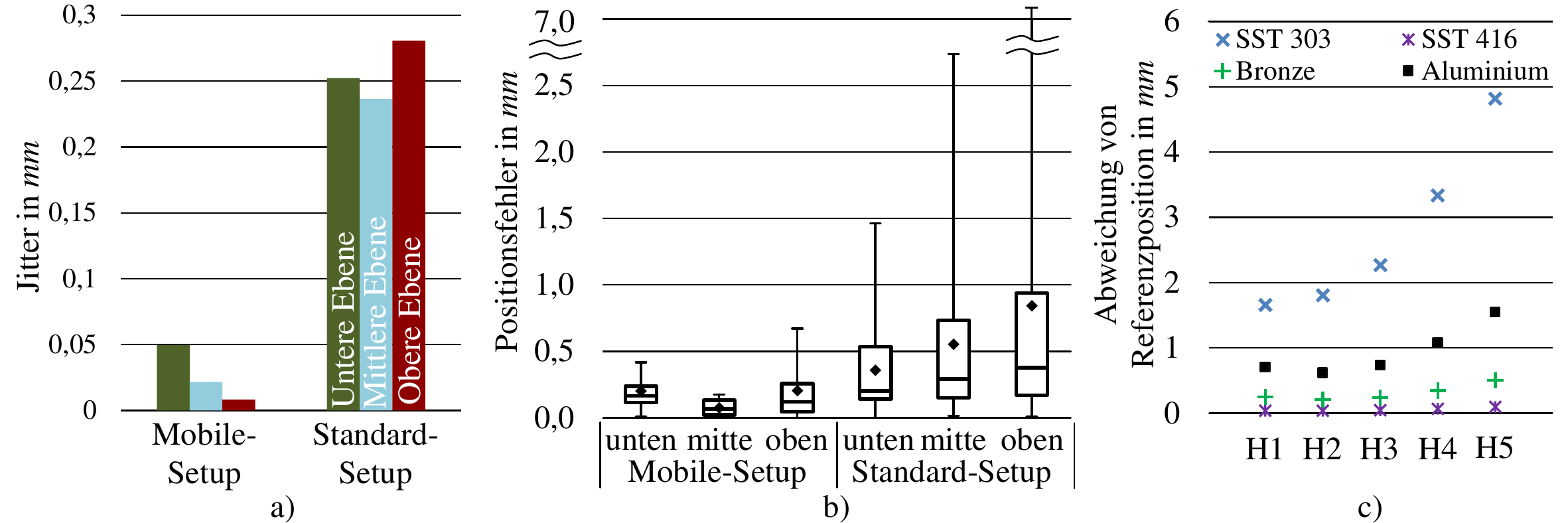}
\vspace*{-0.3cm}
\end{figure}

Die Ergebnisse zu den Positions- und Metallmessungen mit dem \textit{Cable Tool}-Sensor sind in Abb.~\ref{2639-fig:Diagramme} dargestellt. Der mittlere Jitter beträgt im \textit{Mobile-Setup} $0,03$~mm und im \textit{Standard-Setup} $0,26$~mm. Im \textit{Mobile-Setup} beträgt die mittlere Positionsgenauigkeit $0,16$~mm und im \textit{Standard-Setup} $0,59$~mm. Der durch ein Metall verursachte Positionsfehler ist tendenziell umso größer, je näher das \mbox{Metall} am FG platziert ist. Die größte Abweichung von der Referenzposition wird durch Stahl SST 303 hervorgerufen ($4,82$~mm). Bei Stahl SST 416 hingegen fällt die Abweichung am geringsten aus ($0,10$~mm). 

Die Orientierungsfehler betragen im \textit{Mobile-Setup} 0,04° | 0,05° | 0,22° für ROT\_1 |  ROT\_2 | ROT\_3 und im \textit{Standard-Setup} 0,15° | 0,08° | 0,06°.

Die Ergebnisse der Vergleichsmessungen mit dem Mikrosensor fallen in allen Kategorien schlechter aus. Im \textit{Standard-Setup} (mittlere Ebene) liegt der Jitter bei $0,35$~mm und der Positionsfehler bei $0,90$~mm. Der mittlere Orientierungsfehler von ROT\_1, ROT\_2 und ROT\_3 beläuft sich im \textit{Standard-Setup} auf $0,57$°. 

Die Rohdaten aller Messungen und die verwendete Software zur Auswertung der Messdaten haben wir unter \textit{https://osf.io/aphzv/} frei zugänglich veröffentlicht.

\section{Diskussion}

Der \textit{Planar 10-11 FG} wurde mit einem standardisierten Messprotokoll auf seine Genauigkeit und Robustheit geprüft. Die Auswertungsergebnisse belegen, dass der FG trotz des relativ großen Trackingvolumens in störungsfreier Umgebung EM Tracking mit Submillimeter-Genauigkeit ermöglicht.

Im Gegensatz zum \textit{Mobile-Setup} reichen im \textit{Standard-Setup} die Messpositionen bis an den Rand des Trackingvolumens. Da im \textit{Standard-Setup} sowohl der Jitter als auch der Positionsfehler höher ausfallen als im \textit{Mobile-Setup}, ist davon auszugehen, dass die Messgenauigkeit des FG zum Rand des Trackingvolumens hin abnimmt. Dieses Verhalten konnten wir besonders rechts unten an der Eckposition des 5x6-Gitters feststellen (Abb.~\ref{2639-fig:HummelPhantom}). Hier detektierten wir auf der oberen Ebene einen einzelnen Ausreißer mit einem Positionsfehler von $7,16$~mm.

Aus früheren Studien~\cite{2639-07,2639-08,2639-09} ist bekannt, dass EM Trackingsysteme anfällig gegenüber externen Störquellen sind. Auch beim \textit{Planar 10-11 FG} wird der Messfehler durch metallische Störquellen erhöht. Ein quadratischer Zusammenhang zwischen der Entfernung des Metalls zum FG und dem Positionsfehler kann angenommen werden (Abb.~\ref{2639-fig:Diagramme}c). 

Mit dem Mikrosensor erhielten wir ungenauere Messergebnisse im Vergleich zum \textit{Cable Tool}-Sensor. Das betraf besonders die Rotationsmessungen, bei \mbox{denen} einzelne Orientierungsfehler von bis zu 8° registriert werden konnten. Die Messgenauigkeit hängt daher nicht allein vom FG und externen Störquellen ab. Sie wird auch vom verwendeten EM Sensor beeinflusst. Obwohl immer kleinere \mbox{Mikrosensoren} im Submillimeterbereich (0,3~mm) hergestellt werden können~\cite{2639-09}, sollte es daher bei der Planung navigierter medizinischer Interventionen stets zu einer Abwägung kommen zwischen den vielfältigeren Einsatzmöglichkeiten der Mikrosensoren und der für den Eingriff erforderlichen Trackinggenauigkeit.

Vergleichbare Messungen von Maier-Hein et al.~\cite{2639-07} mit dem \textit{Planar FG}, im Folgenden als \textit{Standard FG} bezeichnet, und dem \textit{Compact FG} von NDI ergaben tendenziell schlechtere Genauigkeitsergebnisse. Unter Laborbedingungen beträgt der mittlere Jitter des \textit{Compact FG} 0,05~mm. Er ist damit etwa doppelt so groß wie der Vergleichswert des \textit{Planar 10-11 FG}. Der mittlere Jitter des \textit{Standard FG} fällt mit 0,2~mm etwas kleiner aus als der Jitter des \textit{Planar 10-11 FG}. Sowohl der Positionsfehler als auch der Orientierungsfehler des \textit{Planar 10-11 FG} sind verglichen mit den beiden anderen FG deutlich kleiner (z.B. Positionsfehler im \textit{Mobile-Setup}:  Faktor 3,13 kleiner).

Das in Abb.~\ref{2639-fig:Setup} vorgestellte einfache Setup erleichtert die praktische Anwendbarkeit von EM Trackingsystemen im Klinikalltag. In Kombination mit seiner guten Trackinggenauigkeit könnte der \textit{Planar 10-11 FG} daher aus unserer Sicht dazu beitragen, den Einsatz von IOT in der Klinik zu steigern.
\subsubsection*{Danksagung:}
Das Projekt wurde vom Deutschen Zentrum für Luft- und Raumfahrt (DLR) finanziert (Projekt \textit{OP 4.1}). Herzlichen Dank an die NDI Europe GmbH (Radolfzell, Deutschland) für die Bereitstellung des Feldgenerators. 

\bibliographystyle{bvm2019}

\bibliography{2639}

\begin{thebibliography}{1}

\bibitem{2639-01}
Chartrain AG, Kellner CP, Fargen KM, et~al.
\newblock A review and comparison of three neuronavigation systems for
  minimally invasive intracerebral hemorrhage evacuation.
\newblock J Neurointerv Surg. 2018;10(1):66--74.

\bibitem{2639-02}
Turtulici G, Orlandi D, Corazza A, et~al.
\newblock Percutaneous Radiofrequency Ablation of Benign Thyroid Nodules
  Assisted by a Virtual Needle Tracking System.
\newblock Ultrasound Med Biol. 2014;40(7):1447--1452.

\bibitem{2639-03}
Irugu DVK, Stammberger HR.
\newblock A Note on the Technical Aspects and Evaluation of the Role of
  Navigation System in Endoscopic Endonasal Surgeries.
\newblock Indian J Otolaryngol Head Neck Surg. 2014 Jan;66(1):307--313.

\bibitem{2639-04}
Yoo AJ, Tommy A.
\newblock Thrombectomy in Acute Ischemic Stroke: Challenges to Procedural
  Success.
\newblock J Stroke. 2017;19(2):121--130.

\bibitem{2639-05}
Peters T, editor.
\newblock Image-Guided Interventions.
\newblock Springer--Verlag, Berlin; 2008.

\bibitem{2639-06}
Franz AM, Haidegger T, Birkfellner W, et~al.
\newblock Electromagnetic Tracking in Medicine - A Review of Technology,
  Validation, and Applications.
\newblock IEEE Trans Med Imaging. 2014 Aug;33(8):1702--1725.

\bibitem{2639-07}
Maier-Hein L, Franz AM, Birkfellner W, et~al.
\newblock {{S}tandardized assessment of new electromagnetic field generators in
  an interventional radiology setting}.
\newblock Med Phys. 2012;39(6):3424--3434.

\bibitem{2639-08}
Hummel JB, Bax MR, Figl ML, et~al.
\newblock Design and application of an assessment protocol for electromagnetic
  tracking systems.
\newblock Med Phys. 2005;32(7):2371--2379.

\bibitem{2639-09}
Yaniv Z, Wilson E, Lindisch D, Cleary K.
\newblock Electromagnetic tracking in the clinical environment.
\newblock Med Phys. 2009;36:876--892.

\end{thebibliography}

\end{document}